\documentclass[useAMS,usenatbib]{mn2e}

\usepackage{graphicx}
\usepackage{multirow}

\title[Mutual Shielding Effects in PDRs]
{The Chemical Effects of Mutual Shielding in Photon Dominated Regions}
\author[R.P.Rollins and J.M.C.Rawlings]{R. P. Rollins$^1$\thanks{E--mail:
rpr@star.ucl.ac.uk} and J. M. C. Rawlings$^1$ \\ $^1$Department of Physics and
Astronomy, University College London, Gower Street, London WC1E 6BT, United
Kingdom}

\begin{document}

\maketitle

\begin{abstract}
We investigate the importance of the shielding of chemical photorates by 
molecular hydrogen photodissociation lines and the carbon photoionization 
continuum deep within models of photon dominated regions. In particular, the 
photodissociation of N$_{2}$ and CN are significantly shielded by the
H$_2$ photodissociation line spectrum. We model this by switching off the 
photodissociation channels for these species behind the H{\sc i}$\to$H$_2$ 
transition. We also model the shielding effect of the carbon photoionization 
continuum as an attenuation of the incident radiation field shortwards of 
1102\,\AA. Using recent line and continuum cross section data, we present 
calculations of the direct and cosmic ray induced photorates for a range of 
species, as well as optically thick shielding factors for the carbon 
continuum. Applying these to a time dependent PDR model we see enrichments in 
the abundances of N$_2$, N$_{2}$H$^{+}$, NH$_{3}$ and CN by factors of 
$\sim 3-100$ in the extinction band A$_{v}=2.0$ to A$_{v}=4.0$ for a range of 
environments. While the precise quantitative results of this study are limited by 
the simplicity of our model, they highlight the importance of these mutual 
shielding effects, neither of which has been discussed in recent models.
\end{abstract}

\begin{keywords}
astrochemistry -- ISM: molecules
\end{keywords}

\section{Introduction}
\label{sec:introduction}

Photodissociation or Photon Dominated Regions (PDRs) are pervasive, ubiquitous 
components of the interstellar medium, being the interface between the 
partially ionized/atomic component and the neutral, dense, molecular 
component. Examples of PDRs include the boundaries of molecular clouds 
subjected to the interstellar radiation field, protoplanetary discs irradiated 
by accreting protostar and the interstellar medium of starburst galaxies 
illuminated by clusters of massive star formation.
These interfaces are chemically rich and they often dominate the molecular 
emission from the host regions, due to the relatively high excitation of the 
molecular gas.
Moreover, as earlier, simplistic assumptions about the geometries of these 
regions gives way to more realistic, complex morphologies, it would seem that
a larger proportion of interstellar gas is contained within PDRs.

However, the physical and chemical structure of PDRs is complex and only 
partially described by existing models.
The thermal balance in PDRs is determined by heating by ambient and stellar 
far--ultraviolet radiation ($6<h\nu <13.6$eV) and cooling by atomic/molecular 
line emission and continuum emission by dust (\citealt{HaT:1999};
\citealt{S:2004}). This in turn depends on the chemistry.
The transition from atomic to molecular gas is defined by the radiative 
transfer, coupled to the photochemistry, which entails complex self-- and 
mutual--shielding effects.

It is usually reasonable to assume that H$_2$, CO and its isotopologues are 
the only molecular species that have large enough abundances to result 
in significant absorption of the incident radiation field. 
The primary consequence of this is the increasing self--shielding of the 
photoreaction as the transition becomes more optically thick. In addition,
where there is overlap between the absorption lines of different species, 
mutual shielding will occur.
Also, one must consider the contributions to the UV opacity from the 
carbon ionization continuum (for $\lambda < 1102$\,\AA) and absorption by 
dust. This leads to a range of possible mutual and self--shielding effects as 
listed below:

\begin{enumerate}
\item H$_2$ self--shielding,
\item CO self--shielding,
\item H$_2$, CO and CO isotopologue mutual--shielding,
\item Mutual shielding by coincident Lyman transition lines of H{\sc i} and 
Lyman and Werner transition lines of H$_2$,
\item Shielding of all species by dust absorption, 
\item Shielding of H$_2$ and CO by the carbon ionization continuum, and
\item Shielding of {\it other} species by all of the above processes.
\end{enumerate}

Of these, the cross section for attenuation by the dust continuum is 
approximately constant over the wavelength ranges of individual photoreaction
channels and so is separable from the other shielding terms. 
In practice, the theoretical studies are somewhat limited by the paucity of
laboratory and theoretical data for molecules other than H$_2$ and CO. 

\citet{TaH:1985} developed a PDR model to study the chemistry and 
thermal balance of gas with density $10^3 {\rm cm}^{-3}<n<10^6 {\rm cm}^{-3}$ 
illuminated by far-ultraviolet fluxes $\chi =10^3-10^6$ times more intense 
than the interstellar radiation field.
They identified the presence of several physically distinct regions within 
PDRs, including an H{\sc i}$\to$H$_2$ transition zone at A$_v<1$ and a 
complex, warm C$^+\to$C$\to$CO transition zone at A$_v\sim 3-5$.
Their model, which included carbon continuum self--shielding, yielded C{\sc i} 
column densities of $4.5-16.0\times 10^{18}$cm$^{-2}$ at which levels the 
continuum is very optically thick.
Similarly, \citet{R:1988} developed a photochemical model of the (dust--free)
PDR in the vicinity of a nova, which concentrated on the mutual shielding 
between H$_2$, CO and C{\sc i}.

Subsequent studies, including the PDR comparison/benchmarking study of 
\citet{Rea:2007}, although more sophisticated than the earlier models, 
and employing accurate photoreaction data for CO and H$_2$, essentially 
confirmed the basic structure with the C$^+\to$C$\to$CO transition occurring 
at A$_v\sim 2-5$ for n$\sim 10^3$cm$^{-3}$ and $\chi=10$.
These various models take careful account of the overlap between the CO and 
H$_2$ absorption lines and include the dust continuum, but do not make specific
reference to shielding by the C{\sc i} ionization continuum.

With the availability of reliable wavelength and oscillator strength data 
for the photodissociation of CO, a much more detailed and accurate treatment 
of the self/mutual shielding of H$_2$ and CO has been possible.
\citet{vDaB:1988}
considered the photodissociation of CO and its main isotopologues together 
with H and H$_2$, including self-- and mutual--shielding effects.
They also included shielding by dust and referred to the attenuation by the 
carbon continuum.
As shown in that paper (e.g. see Fig.~4), when the CO, H{\sc i} and H$_2$ 
absorption lines become very optically thick, they can effectively blanket a 
significant fraction of the continuum in the 
$912$\,\AA\,$<\lambda<1079$\,\AA\, wavelength range, 
with implications for the photodissociation rates of other species. 

\citet{Vea:2009} 
included self--shielding, mutual--shielding and shielding by H and H$_2$ to
calculate the photodissociation rates of CO and its isotopologues, including 
rarer variants.
For these species only, the photodissociation rate coefficient can be written 
as:

\begin{equation}
\label{eq:rate}
k = \chi k_0\Theta\left[ N(H_2),N(CO),N(H),\phi\right] \exp (-\gamma A_v),
\end{equation} 

where $k_0$ is the unattenuated rate, $\Theta$ is the shielding function, 
$\exp (-\gamma A_v)$ is the (separable) dust extinction term,
and $\phi$ represents other factors, such as the Doppler widths of the 
absorption lines, the excitation temperatures for CO and H$_2$, and ratios of 
the column densities of the CO isotopologues.
As with previous studies the line and continuum attenuation components were
treated as separable.
However, as pointed out by \citet{vD:1988} the A$_v$ dependence does 
{\em not} include the attenuation at $\lambda<1102$\,\AA\, by H$_2$ 
line/continuum and C{\sc i} continuum absorption which will be particularly 
important for species like CN and N$_2$, which photodissociate at the shortest 
wavelengths.  

Thus, the earliest studies of PDRs (e.g. \citealt{TaH:1985}) -- 
although lacking the details of more recent models -- attempted to deal with 
most of the shielding effects listed above.
In subsequent studies all discussion of the role of the carbon 
ionization continuum is seemingly absent. This is presumably on the assumption 
-- which may or may not be valid -- that dust extinction dominates.
However, the cross section for carbon photoionization is 
$1.6\times 10^{-17}$cm$^2$.
It is fairly reasonable to assume that column densities of atomic carbon in 
excess of 10$^{17}$cm$^{-2}$ will pertain deep within PDRs 
(e.g. Fig~11 of \citealt{Vea:2009}) and so this opacity source must be 
considered.

In addition, the more recent PDR models have tended to concentrate on the 
carbon budget, while the effects of mutual shielding by H$_2$ and CO on 
{\em other} molecules have been largely overlooked.
In this study we consider two of these effects; the mutual shielding by:
\begin{enumerate}
\item molecular hydrogen lines, and
\item the carbon ionization continuum.
\end{enumerate}
Critically, there is extensive overlap of both of these with the 
photodissociation cross sections of CN and N$_2$. 
The mutual shielding effects therefore have the potential 
to promote nitrogen-based chemistry at extinctions much shallower than 
currently seen in simulations. This paper presents shielding 
factors for a range of species due to these effects and investigates their 
influence on the chemical structure.

Our calculation of the degree of mutual shielding by H$_{2}$ dissociation and 
carbon ionization are presented in Section \ref{sec:photoreactions}. These 
shielding factors are then implemented in a simple PDR model, described in 
Section \ref{sec:model} with results presented in Section \ref{sec:results}. 
We close with a discussion of our results and concluding remarks in Section 
\ref{sec:conclusions}.

\section{Photoreaction Rates}
\label{sec:photoreactions}

In PDRs, the shielding of the incident radiation field by molecules is
typically a combination of line and continuum processes leading to a complex 
radiative transfer problem. Most models include self shielding by CO and 
H$_{2}$ by adding multiplicative shielding factors to their photorates as 
described above.
We extend this procedure to investigate mutual shielding by the 
H$_2$ line and C{\sc i} ionization continuum absorption.
In this section we detail our method for calculating photoreaction rates and 
shielding factors for these species.

\subsection{Direct Photorates}
\label{sec:photorates}

\begin{table*}
\caption[photoionization]{Photoionization Rates, Carbon Continuum Shielding 
Factors, Cosmic Ray Induced Photoreaction Efficiencies and Cross Section Data 
References; \citealt{vD:1988} 
(http://home.strw.leidenuniv.nl/$\sim$ewine/photo/, vD88), \citealt{B:1962} (B62), 
\citealt{B:1977} (B77), \citealt{BD:1977} (BD77) and \citealt{Rea:1993} (RDB93).}
\centering
\begin{tabular}{|c|c|c|c|c|}
\hline
\multirow{2}{*}{Species}&\multirow{2}{*}{Photorate / s$^{-1}$}&Maximal Carbon&CR Induced Photoreaction&\multirow{2}{*}{Reference}\\
&&Shielding Factor, S$_i$&Efficiency, p$_M$\\
\hline
C&$3.15\times10^{-10}$&0.000&264.8&vD88\\
C$^-$&$3.31\times10^{-08}$&0.994&1037.8&B62\\
CH&$7.63\times10^{-10}$&0.142&547.7&vD88\\
CH$_2$&$3.58\times10^{-10}$&0.449&300.0&B77\\
CH$_3$&$4.93\times10^{-10}$&0.600&509.5&BD77\\
NH&$1.50\times10^{-11}$&0.000&22.0&B77\\
NH$_2$&$1.73\times10^{-10}$&0.000&149.3&RDB93\\
NH$_3$&$2.82\times10^{-10}$&0.298&230.5&vD88\\
O$_2$&$7.65\times10^{-11}$&0.000&57.7&vD88\\
OH&$2.00\times10^{-11}$&0.000&26.6&B77\\
H$_2$O&$3.44\times10^{-11}$&0.000&24.5&vD88\\
Na&$1.54\times10^{-11}$&0.829&9.1&vD88\\
HCO&$4.81\times10^{-10}$&0.590&499.2&RDB93\\
H$_2$CO&$4.82\times10^{-10}$&0.215&379.5&vD88\\
NO&$2.61\times10^{-10}$&0.328&218.9&vD88\\
S&$6.06\times10^{-10}$&0.343&480.2&vD88\\
H$_2$S&$7.40\times10^{-10}$&0.344&594.4&vD88\\
\hline
\end{tabular}
\label{table:photoionization}
\end{table*}

\begin{table*}
\caption[photodissociation]{Photodissociation Rates, Carbon Continuum 
Shielding Factors, Cosmic Ray Induced Photoreaction Efficiencies and Cross 
Section Data References; \citealt{vD:1988} 
(http://home.strw.leidenuniv.nl/ewine/photo/, vD88), \citealt{KB:1978} (KB78) 
and \citealt{C:1972} (C72).}
\centering
\begin{tabular}{|c|c|c|c|c|}
\hline
\multirow{2}{*}{Species}&\multirow{2}{*}{Photorate / s$^{-1}$}&Maximal Carbon&CR Induced Photoreaction&\multirow{2}{*}{Reference}\\
&&Shielding Factor, S$_i$&Efficiency, p$_M$\\
\hline
H$_2^+$ $\to$ H$^+$ + H&$5.73\times10^{-10}$&0.778&427.7&vD88\\
H$_3^+$ $\to$ H$_2^+$ + H&$2.16\times10^{-12}$&0.000&0.15&KB78\\
H$_3^+$ $\to$ H$_2$ + H$^+$&$2.16\times10^{-12}$&0.000&0.15&KB78\\
CO $\to$ O + C&$2.59\times10^{-10}$&0.000&105.0, see text&vD88\\
CO$^+$ $\to$ C$^+$ + O&$1.01\times10^{-10}$&0.954&45.7&vD88\\
CH $\to$ C + H&$9.19\times10^{-10}$&0.975&465.4&vD88\\
CH$^+$ $\to$ C + H$^+$&$3.27\times10^{-10}$&0.036&254.7&vD88\\
CH$_2$ $\to$ CH + H&$5.81\times10^{-10}$&1.000&200.4&vD88\\
CH$_2^+$ $\to$ CH$^+$ + H&$4.53\times10^{-11}$&0.461&66.1&vD88\\
CH$_2^+$ $\to$ CH + H$^+$&$4.53\times10^{-11}$&0.461&66.1&vD88\\
CH$_2^+$ $\to$ C$^+$ + H$_2$&$4.53\times10^{-11}$&0.461&66.1&vD88\\
CH$_3$ $\to$ CH$_2$ + H&$1.38\times10^{-10}$&1.000&76.1&vD88\\
CH$_3$ $\to$ CH + H$_2$&$1.38\times10^{-10}$&1.000&76.1&vD88\\
CH$_4$ $\to$ CH$_3$ + H&$1.89\times10^{-10}$&0.703&182.8&vD88\\
CH$_4$ $\to$ CH$_2$ + H$_2$&$8.42\times10^{-10}$&0.703&182.8&vD88\\
CH$_4$ $\to$ CH + H$_2$ + H&$1.89\times10^{-10}$&0.703&182.8&vD88\\
CH$_4^+$ $\to$ CH$_2^+$ + H$_2$&$2.23\times10^{-10}$&0.405&139.3&vD88\\
CH$_4^+$ $\to$ CH$_3^+$ + H&$5.23\times10^{-11}$&0.405&139.3&vD88\\
NH $\to$ N + H&$5.02\times10^{-10}$&0.794&219.2&vD88\\
NH$_2$ $\to$ NH + H&$7.45\times10^{-10}$&0.783&152.8&vD88\\
NH$_3$ $\to$ NH + H$_2$&$4.83\times10^{-10}$&0.780&159.2&vD88\\
NH$_3$ $\to$ NH$_2$ + H&$6.82\times10^{-10}$&0.891&148.1&vD88\\
N$_2$ $\to$ N + N&$2.28\times10^{-10}$&0.000&184.3&vD88, C72, see text\\
O$_2$ $\to$ O + O&$7.90\times10^{-10}$&0.890&361.3&vD88\\
O$_2^+$ $\to$ O$^+$ + O&$3.46\times10^{-11}$&1.000&3.0&vD88\\
OH $\to$ O + H&$3.76\times10^{-10}$&0.752&279.3&vD88\\
OH$^+$ $\to$ O$^+$ + H&$1.30\times10^{-11}$&0.008&10.7&vD88\\
H$_2$O $\to$ OH + H&$7.54\times10^{-10}$&0.746&527.0&vD88\\
HCO $\to$ CO + H&$1.11\times10^{-09}$&1.000&319.8&vD88\\
HCO$^+$ $\to$ CO$^+$ + H&$5.39\times10^{-12}$&0.000&0.0&vD88\\
H$_2$CO $\to$ CO + H + H&$5.79\times10^{-10}$&0.844&132.1&vD88\\
H$_2$CO $\to$ CO + H$_2$&$5.79\times10^{-10}$&0.844&132.1&vD88\\
H$_2$CO $\to$ HCO$^+$ + H + e$^-$&$1.16\times10^{-11}$&0.844&132.1&vD88\\
CO$_2$ $\to$ CO + O&$8.81\times10^{-10}$&0.329&643.8&vD88\\
CN $\to$ N + C&$2.92\times10^{-10}$&0.018&229.2&vD88\\
HCN $\to$ CN + H&$1.56\times10^{-09}$&0.677&436.9&vD88\\
HNC $\to$ CN + H&$1.56\times10^{-09}$&0.677&436.9&vD88, see text\\
NO $\to$ O + N&$4.73\times10^{-10}$&0.737&163.5&vD88\\
HS $\to$ S + H&$9.79\times10^{-10}$&0.947&271.9&vD88\\
HS$^+$ $\to$ S$^+$ + H&$2.60\times10^{-10}$&0.343&171.9&vD88\\
H$_2$S $\to$ HS + H&$1.55\times10^{-09}$&0.845&412.6&vD88\\
H$_2$S $\to$ S + H$_2$&$1.55\times10^{-09}$&0.845&412.6&vD88\\
CS $\to$ S + C&$9.75\times10^{-10}$&0.973&1570.0&vD88\\
\hline
\end{tabular}
\label{table:photodissociation}
\end{table*}

It is usually assumed that the line and continuum contributions to the 
photorates are separable \citep{vD:1988}.
Unshielded rate coefficients, $k_{pd}$, can be calculated using the following
equations for line and continuum photoprocesses (e.g. \citealt{vD:1988}):

\begin{equation}
\label{eq:continuum}
k_{pd}^{cont}=\int_{912\AA}^{\infty}\sigma(\lambda)I(\lambda) \, 
d\lambda \, s^{-1},
\end{equation}

\begin{equation}
\label{eq:line}
k_{pd}^{line}=\frac{\pi e^{2}}{mc^{2}}\lambda_{ul}^{2}f_{ul}\eta_{u}x_{l}
I(\lambda_{ul}) \, s^{-1},
\end{equation}

where $I$ is the intensity of the radiation field, $\sigma$ is the continuum 
cross section and the line transition is at a wavelength $\lambda_{ul}$ from a 
lower level $l$ to an upper level $u$ with oscillator strength $f_{ul}$, 
dissociation efficiency $\eta_{u}$ for 
the upper level and occupation fraction $x_{l}$ for the lower level. 
The lower limit of the integral (912\,\AA) is the Lyman cutoff wavelength for 
the unshielded interstellar radiation field.
From these expressions we have (re-)calculated the unshielded interstellar 
photorates using cross section and oscillator strength data primarily 
from \citet{vD:1988}, referring to other sources (e.g. \citealt{Rea:1993}) 
for species and reaction channels not covered in their database. 
Where we have one set of cross section data for a molecule but 
multiple dissociation channels are listed in the UMIST Database for 
Astrochemistry (UMIST06, \citealt{Wea:2007}), our calculated rate is divided 
between the channels in the same ratio as the original UMIST06 rates.
We note that the high resolution N$_2$ dissociation line data from 
\citet{C:1972} has been re--binned to a lower wavelength resolution in
the database of \citet{vD:1988}. As the representative wavelengths, we therefore 
elected to use the mid-points of the dissociation wavelength intervals from 
\citet{C:1972}.
Since no cross section data is available for HNC we assign the same values as 
derived for HCN.
For the radiation field we adopt the standard interstellar radiation 
field of \citet{D:1978}, with the extension to $\lambda >2000$\,\AA\, given by 
\citet{vDaB:1982}. These photorates are presented in Tables 
\ref{table:photoionization} and \ref{table:photodissociation}.

The newly calculated direct photorates are in agreement with the values in 
UMIST06 to within a factor of two for all but 10 of the 61 reactions for which 
we had data available. Four show rates differing by more than one order of 
magnitude; our calculations for the dissociation of H$_{3}^{+}$ to either 
H$_{2}^{+}$ + H or H$_{2}$ + H$^{+}$ are each larger by a factor of $\sim400$,
while the ionization of OH is larger by a factor of $\sim 12$ and the 
dissociation of OH$^+$ to O$^+$ + H is larger by a factor of $\sim19$. 

\subsection{Shielding of N$_2$ and CN by H$_2$}
\label{sec:h2}

The photodissociation of H$_{2}$ is effected by the absorption of Lyman and 
Werner band photons followed by decay into the dissociation continuum 
\citep{sea:1967}.
To investigate the possible effects of mutual shielding by these lines we 
note that the range of wavelengths for H$_2$ dissociation ($844.8-1108.5$\,\AA)
fully overlaps that for the dissociation lines of N$_2$ ($914-980$\,\AA, 
\citealt{vD:1988}, \citealt{C:1972}) and nearly the whole range for CN photodissociation ($912-1133$\,\AA, peaking at $940$\,\AA).
We are investigating the photochemistry well within a PDR (A$_v>2$) and the 
H$_2$ lines will be extremely optically thick, to the extent that (a) the
overlap with individual N$_2$ lines will be extensive and (b) there will
be significant blanketing of the Lyman/Werner continuum.
To include these effects we therefore make the extreme assumption that 
N$_2$ and CN are both {\em fully} shielded and that the effective rates of 
photodissociation can be set to zero. These assumptions are clearly very 
simplistic and will need addressing using more sophisticated radiative 
transfer techniques in a future study.

\subsection{Shielding by C{\sc i}}
\label{sec:ci}

To account for shielding by the C{\sc i} ionization continuum we take a more
detailed and accurate approach.
First of all, considering the situation when the continuum is very optically 
thick, the photoreaction rates are re-calculated using 
equations \ref{eq:continuum} and \ref{eq:line}, but using a lower limit of 
$1102$\,\AA\, for the integral and excluding line contributions for 
$\lambda<1102$\,\AA. This implicitly assumes that the radiation field 
intensity is completely attenuated at these wavelengths shortwards of the 
C{\sc i} ionization potential; I($\lambda< 1102$\,\AA)=0.
In Tables \ref{table:photoionization} and \ref{table:photodissociation} we 
give the resulting shielding factors ($S_i$), which represent the ratios of the
photoreaction rates without and with the contributions from the 
$912<\lambda <1102$\,\AA\, wavelength range. Thus, in conditions where the 
carbon ionization continuum is very optically thick, the rate coefficient is 
just $S_i$ times the unshielded value.
These ratios are calculated for all species/channels where cross section 
data is known. 

We note that the photoionization cross section of atomic carbon is 
$\sigma_C= 1.6\times10^{-17}$ cm$^{2}$ and is approximately independent of 
wavelength for $\lambda< 1102$\,\AA\, \citep{vD:1988}. 
C{\sc i} column densities of up to the order of $10^{19}$ cm$^{-2}$ 
\citep{TaH:1985} can be expected in PDRs, leading to a wide range of 
potential optical depths. Taking the first approximation that attenuation by 
dust and molecules are separable and the fact that the cross section is 
uniform with wavelength simplifies the radiative transfer to give shielding 
functions, $\Theta_i$, for each photoreaction $i$ (analogous to those used 
for CO):

\begin{equation}
\label{eq:shielding}
\Theta_{i}(N_C) = S_{i}+(1-S_{i}){\rm e}^{-\sigma_{C} N_{C}},
\end{equation}

where $S_i$ is the relevant shielding factor for the optically thick limit, 
given in Tables \ref{table:photoionization} and \ref{table:photodissociation}, 
and N$_{C}$ is the carbon (C{\sc i}) column density. Strictly speaking, the 
Av-dependence (due to continuum absorption by dust; $\gamma$ in equation 
\ref{eq:rate}) will also depend on the C{\sc i} continuum opacity. However, this 
is a complicating second-order effect and we follow the practice of previous 
studies by decoupling the shielding by C{\sc i} from that by dust absorption, as 
in equation \ref{eq:rate}.

\subsection{Cosmic Ray Induced Photoreactions}
\label{sec:cosmicRays}

At extinctions greater than a few magnitudes, or if any of the various 
shielding effects described above become important, cosmic ray induced 
photodissociation and photoionization processes are significant and must be 
included in the models.
The radiation field is generated by secondary electrons produced in the 
cosmic ray ionization of H$_2$ \citep{PaT:1983}. 
The calculation of these rates is as described in \citet{Sea:1987} and 
\citet{Gea:1989}:

On the assumption that the total absorption cross section is dominated by 
dust (rather than the molecular component, which in normal circumstances is 
true, except for H$_2$ and CO) then the photorates are given by:

\begin{equation}
\label{eq:crrate}
\rm R_i = \frac{\zeta n(i)}{(1-\omega)}p_i cm^{-3}s^{-1},
\end{equation}

where $\zeta$ is the cosmic ray ionization rate, n(i) is the abundance of 
species $i$, and $\omega$ is the grain albedo. The cosmic ray induced 
photoreaction efficiency, $p_i$, is given by:

\begin{equation}
\label{eq:efficiency}
p_i = \int \frac{\sigma_i(\nu)P(\nu)}{2\sigma_g}d\nu,
\end{equation}

where $\sigma_i$ is the photoreaction cross section and $\sigma_g$ is the 
grain extinction cross section per H--nucleon. 
The factor of 2 in the denominator takes into account the fact that the
definition of $\sigma_g$ is per H--nucleon, whereas $\zeta$ is defined per 
H$_2$ molecule \citep{Wea:2007}.
We have re-calculated the values of $p_i$ using the same cross section 
data as in Section \ref{sec:photorates} and the (high resolution) cosmic ray 
induced H$_2$ emission spectrum, $P(\nu)$ (Gredel, personal communication, as 
depicted in Figure 1 of \citealt{Gea:1989}). 
This spectrum has been normalised to take account of the various transition 
probabilities and excitations per cosmic ray ionization. Thus it includes 
contributions from excitations to the
$B ^{1}\Sigma_u^+$, $B^{\prime 1}\Sigma_u^+$, $B^{\prime\prime 1}\Sigma_u^+$,
$C ^{1}\Pi_u$, $D ^{1}\Pi_u$, and $D^{\prime 1}\Pi_u$ 
Rydberg states and to the valence $E, F ^1\Sigma_g^+$ and $a ^3\Sigma_g^+$ 
states. Also included are excitations into the repulsive $b ^3\Sigma_u^+$ 
state and the vibrational levels of the ground state ($X ^1\Sigma_g^+$), 
together with cascades to the $B ^1\Sigma_u^+$ state.
The ratio of the H$_2$ populations in J=0 to J=1 is taken to be $1:3$. 
Results of these calculations are also given in 
Tables \ref{table:photoionization} and \ref{table:photodissociation}.
Note that we have not re-calculated the cosmic ray induced rate for the 
photodissociation of CO, for which multiple line overlap occurs and the values 
given in the UMIST06 database are reasonably accurate.

For 24 of the 61 reaction channels, cosmic ray induced photoreaction
efficiencies are not specified in the UMIST06 database. Of the remaining 36 
(excluding CO), all but 15 agree to within a factor of 2. Relative to the
UMIST06 database we found the rates for the ionization of NH, the dissociation
of CN and the dissociation of H$_{2}$CO to H$_{2}$ + CO are all lower by
factors of $\sim$ 12, 25 and 10 respectively.

\section{The Chemical Model}
\label{sec:model}

\begin{table}
\caption[models]{Models}
\begin{tabular}{|c|c|c|c|}
\hline
Model & Definition    &$\chi$  &n / cm$^{-3}$\\ 
\hline
F1        &Standard       &10        &10$^{3}$    \\
F2        &Bright      &10$^{5}$  & 10$^{3}$   \\
F3        &Dense       &10        &10$^{5.5}$  \\
\hline
\end{tabular}
\label{table:models}
\end{table}

\begin{table}
\caption[Physical Parameters]{Physical Parameters}
\begin{tabular}{|c|c|}
\hline
Temperature, T&$50$ K \\
CR Ionization Rate, $\zeta$&$5\times10^{-17}$ s$^{-1}$ \\
Grain Albedo, $\omega$&$0.5$ \\
X(He)& $10^{-1}$\\
X(C)& $10^{-4}$\\
X(O)& $3\times10^{-4}$\\
X(N)& $8\times10^{-5}$\\
X(Na)& $10^{-6}$\\
X(S)& $10^{-6}$\\
\hline
\end{tabular}
\label{table:parameters}
\end{table}

\begin{table}
\caption[models]{Chemical species used in the model}
\begin{tabular}{|c|}
\hline
Chemical Species \\ 
\hline
H, H$^+$, H$^-$, H$_2^+$, H$_3^+$ \\
He, He$^+$, Na, Na$^+$, e$^-$ \\
C, C$^+$, C$^-$, CO, CO$^+$, CH, CH$^+$, CH$_2$, CH$_2^+$ \\
CH$_3$, CH$_3^+$, CH$_4$, CH$_4^+$, CH$_5^+$ \\
N, N$^+$, NH, NH$^+$, NH$_2$, NH$_2^+$, NH$_3$, NH$_3^+$, NH$_4^+$, N$_2$, 
N$_2^+$, N$_2$H$^+$ \\
O, O$^+$, O$_2$, O$_2^+$, OH, OH$^+$, H$_2$O, H$_2$O$^+$, H$_3$O$^+$ \\
HCO, HCO$^+$, H$_2$CO, H$_2$CO$^+$, CO$_2$, CO$_2^+$, HCO$_2^+$ \\
CN, CN$^+$, HCN, HCN$^+$, HNC \\
NO, NO$^+$, HNO, HNO$^+$, HCNH$^+$, H$_2$NC$^+$, HNCO$^+$, H$_2$NO$^+$ \\
S, S$^+$, HS, HS$^+$, H$_2$S, H$_2$S$^+$, H$_3$S$^+$ \\
CS, CS$^+$, C$_2$S, C$_2$S$^+$, HC$_2$S$^+$, HCS, HCS$^+$, H$_2$CS$^+$ \\
\hline
\end{tabular}
\label{table:species}
\end{table}

To investigate the impact of the shielding mechanisms discussed above on 
chemical abundances, we implement them in a time-- and depth--(A$_{v}$) 
dependant chemical model. Since we are not attempting to replicate any 
specific astrophysical environment, we adopt physical and chemical parameters 
similar to the PDR code benchmarking effort of \citet{Rea:2007}. These are 
summarised in Tables \ref{table:models} and \ref{table:parameters}. In 
particular, we do not attempt to solve the equations of thermal balance and 
the temperature and density (T and n) are fixed. We also assume that the PDR 
is dynamically static so that the extinction, A$_{v}$, is only a function of 
position. The incident radiation field used is the standard interstellar 
radiation field as described above, scaled by the factor $\chi$.

Our chemistry consists of 7 elements in 81 gas phase species 
(Table \ref{table:species}) linked by 1129 reactions. For the sake of 
simplicity, we do not include any gas--grain chemical reactions.
In any case, at the assumed temperature of 50K, gas--grain interactions 
will have negligible significance. The elemental helium, carbon and 
oxygen abundances are the same as in \citet{Rea:2007} while for nitrogen, 
sodium and sulphur, which were absent from their models, we take abundances 
from \citet{Sea:1992} and \citet{Aea:2005}. Reaction data and rate 
coefficients are taken from the UMIST06 database except where we calculated 
updated photorates as described in Section \ref{sec:photoreactions}. 
We assume a fixed cosmic ray ionization rate of $5.0\times10^{17}$ s$^{-1}$ and
a grain albedo of 0.5. The cosmic ray induced photodissociation rate for CO is 
modified by a temperature dependant factor of (T/300K)$^{1.17}$ 
\citep{Wea:2007} yielding a factor of eight reduction in the rate at 50K. 
For species with no cross section data available we use a default cosmic ray 
induced photoreaction efficiency of 200; broadly representative of the average 
value for other species.

The incorporation of the various shielding effects uses a number of 
simplifying approximations to allow for a focused study on the effects of 
H$_{2}$ and C{\sc i} shielding. As in other studies we treat the attenuation 
by dust of the radiation field as separable from other processes and  
described by the usual exponential dependences on extinction of the 
photorates. Strictly speaking, these dependences should be modified to take 
account of the C{\sc i} absorption etc., but such variations are relatively 
minor as compared to the other effects reported in this paper.

For standard PDR models (e.g. model F1 of \citealt{Rea:2007}),
the transition from atomic to molecular hydrogen occurs at A$_{v} \sim 0.1$,
the C{\sc ii}$\to$C{\sc i} transition is at A$_{v} \sim 1.0$ and carbon is 
mostly converted to CO for A$_{v} > 2.0$. 
Deeper than this we can assume sufficient columns of H$_{2}$ and CO have 
formed that they are completely self shielded and the effective 
photodissociation rates for these species are zero. 
For the same reason, when we investigate the effects of the inclusion of 
mutual shielding by molecular hydrogen lines, we take the extreme case 
scenario described in Section \ref{sec:photoreactions} and set the 
photodissociation rates of N$_{2}$ and CN to zero. We therefore restrict our 
investigation to A$_{v}>2.0$ where shielding by 
H$_{2}$ and CO take this simple form and do not attempt to consider the 
details of the PDR itself, which we defer to a later study.
Within our model we can therefore crudely switch the H$_{2}$ line mutual
shielding on or off and assign a fixed carbon column density across the
whole extinction range so as to investigate the
interplay between the two mechanisms and their effect on the chemistry.

\begin{figure*}
\includegraphics[scale=0.75]{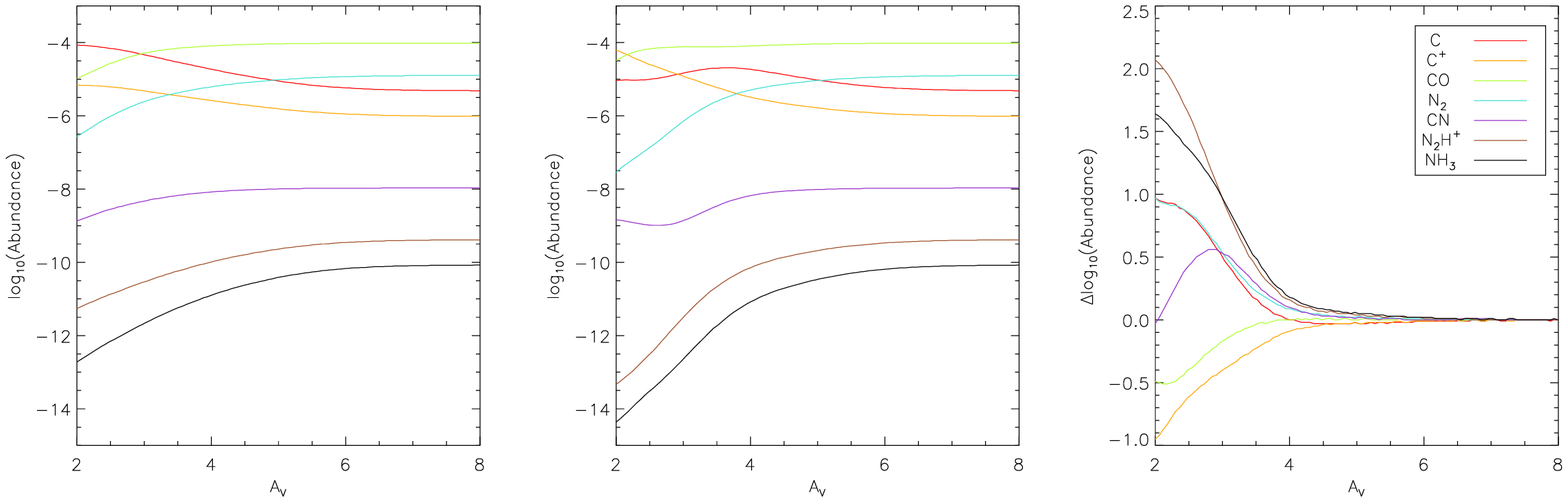}
\includegraphics[scale=0.75]{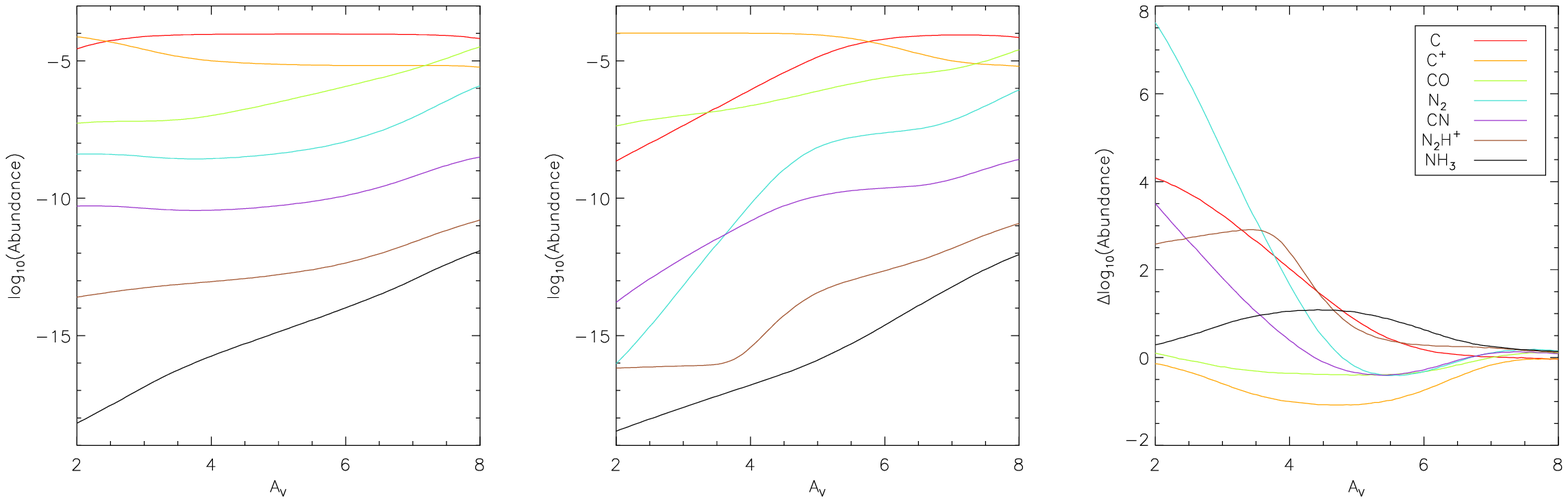}
\includegraphics[scale=0.75]{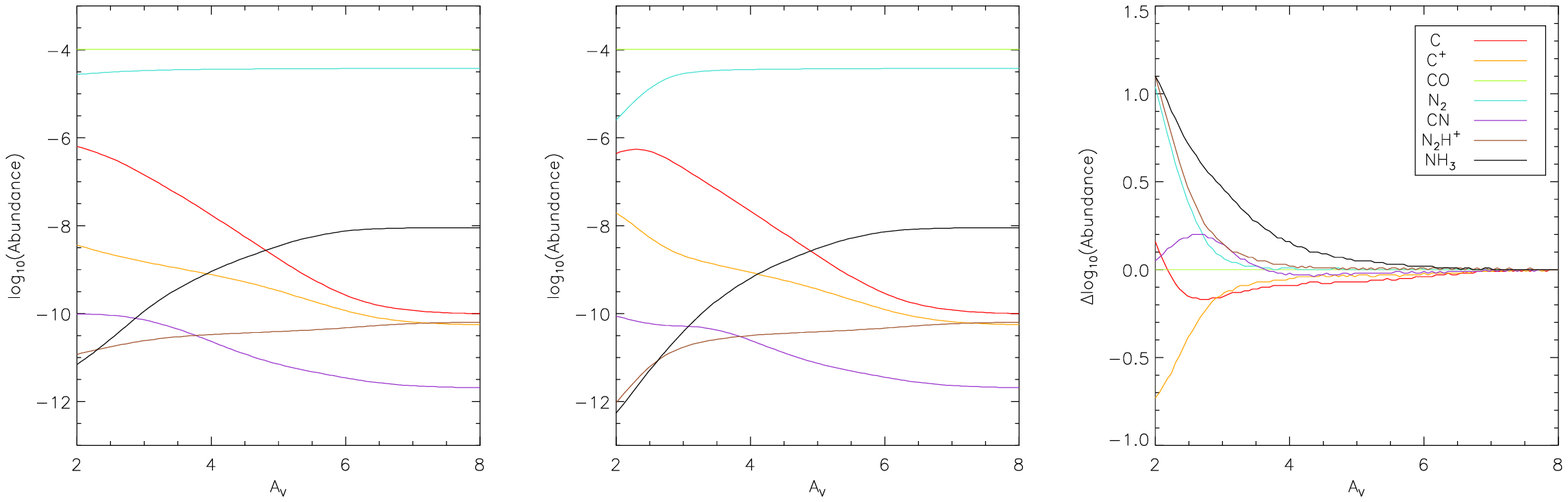}
\caption{Abundances for models with H$_{2}$ mutual shielding and $\tau(C)=10$ 
(left), without H$_{2}$ mutual shielding and $\tau(C)=0$ (centre) and relative 
enrichments between the two (right). Top: Standard (F1), Middle: Bright (F2), 
Bottom: Dense (F3).}
\label{figure:models}
\end{figure*}

\begin{figure*}
\includegraphics[scale=0.75]{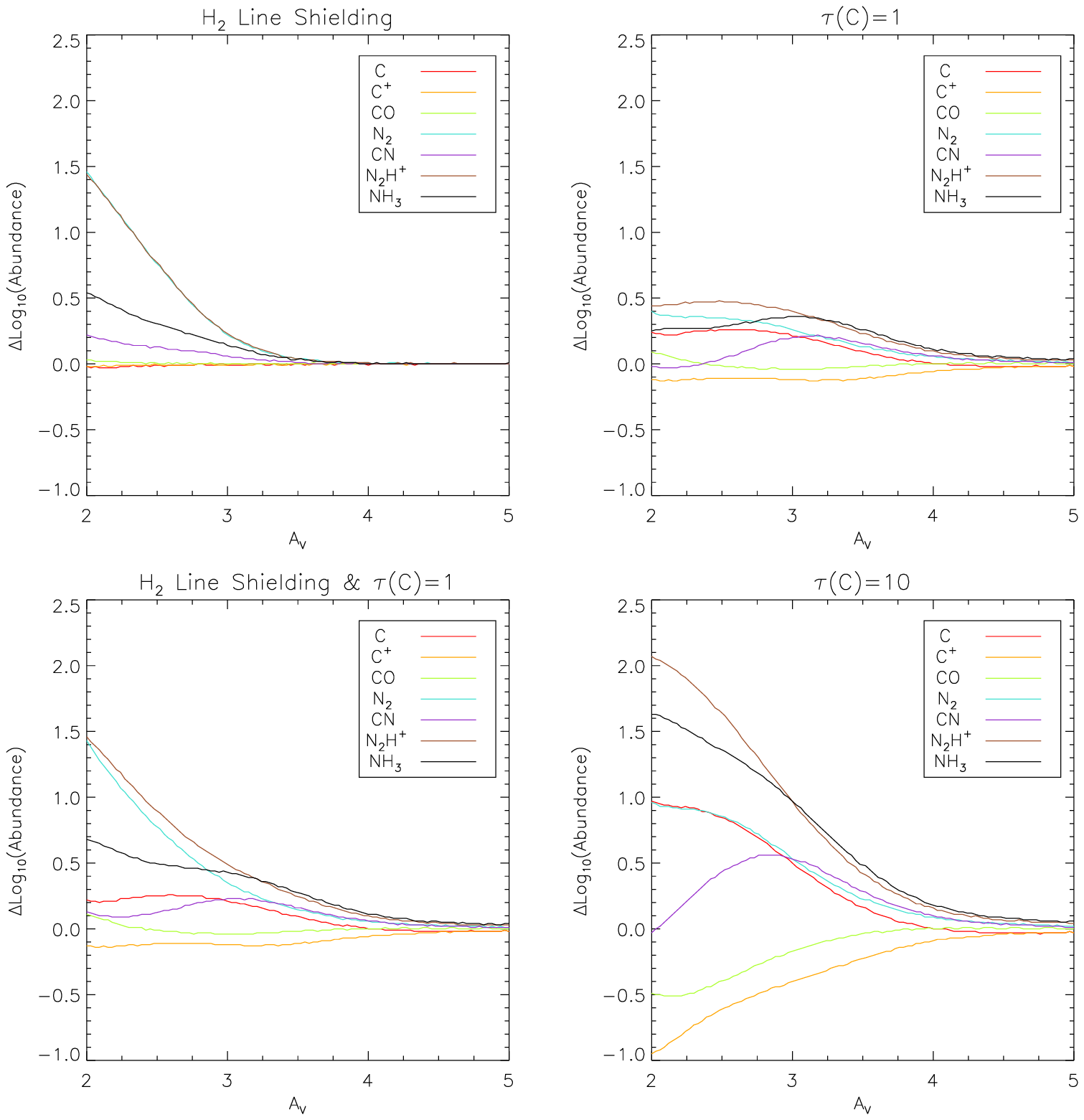}
\caption{Enrichments from model F1 for a range of H$_2$ and C{\sc i} 
shielding regimes, all relative to the case where there is no H$_{2}$ 
mutual shielding and $\tau(C)=0$. 
Note that in the first panel, the curves for N$_{2}$ and 
N$_{2}$H$^{+}$ coincide.}
\label{figure:shieldings}
\end{figure*}

\section{Results}
\label{sec:results}

Examples of the key results obtained from the model are given in 
Figures \ref{figure:models} and \ref{figure:shieldings}. 
Our model was tested against the results for C$^+$, C and CO abundances
in model F1 of \citet{Rea:2007}. We are able to reproduce abundances of
C$^+$, C and CO at positions well within the PDR, while observing typical 
sensitivities to the C/O ratio, the temperature dependence of the cosmic ray 
induced CO photodissociation rate and the chemical age. For a direct 
comparison we have evolved the chemistry to equilibrium. This typically takes 
$\sim 3\times 10^6$ years, although we integrate to 100\,Myr so as to be 
consistent with their calculations.

The figures give a clear comparison of the chemical structure that is obtained 
with and without the implementation of H$_{2}$ mutual shielding and carbon 
continuum shielding.
We present abundances for observable molecules with 
significant abundances and showing abundance variations of factors greater 
than 10 as a result of these two mechanisms. The results are given
in Figure \ref{figure:models} for the models F1, F2 and F3. We 
also show abundances for C$^{+}$, C and CO (PDR carbon transition species)
as well as N$_{2}$ and CN (species shielded by H$_{2}$) for reference. 
Probably the most important effect on the chemistry is the shielding of N$_2$, 
for which the photodissociation is completely suppressed by both mechanisms. 
Thus, very significant enrichments are visible in all three environments and 
particularly in the range of extinctions A$_{v}\sim2-4$. This abundance 
excess obviously drives a more vigorous nitrogen chemistry so that the 
abundances of other nitrogen-bearing species, most notably N$_{2}$H$^{+}$, 
NH$_{3}$ and CN, are significantly enhanced.
The photoionization and photodissociation of NH$_3$ are both partially 
shielded by the carbon continuum, while CN is fully shielded by molecular 
hydrogen. As is usually the case for molecular ions, photodissociation of 
N$_{2}$H$^{+}$ is not significant and had been excluded from our chemical 
network. The presence of significant column densities of nitrogen-bearing 
molecules at relatively low extinctions is an important result.
However, it is also interesting to note that so {\em few} species are 
significantly enhanced by the shielding processes.   
For example, one might have expected to see large increases in 
HCN and HNC, but while their enrichments are tightly correlated they are still 
limited to factors of less than 10. Similarly, despite the strong 
enrichments in NH$_{3}$, the simpler hydrides NH and NH$_{2}$ show correlated 
enrichments but of less than an order of magnitude.

One issue with these particular models is the assumption that the C{\sc i} 
continuum is very optically thick; with $\tau(C)=10$ at A$_{v}=2$. 
Using a typical relation for $A_{v}=6.289\times10^{-22}N_{H_{total}}$ 
\citep{Rea:2007} depending on the total hydrogen column density 
$N_{H_{total}}$ allows us to approximate the atomic carbon column density. For 
model F1, a column of roughly 10$^{16}$ cm$^{-2}$ is obtained at A$_{v}=2$,
rising to $\sim 10^{17}$ cm$^{-2}$ at A$_{v}=4$, implying a C{\sc i} optical 
depth of order unity, but somewhat less than $\tau(C)=10$. The
other models yield lower values, with the bright model not achieving a column 
of 10$^{17}$ cm$^{-2}$ until A$_{v}\sim 6$ and the dense model only 
achieving a maximum column of roughly $\sim 10^{15}$ cm$^{-2}$. However, the 
A$_{v}$--N$_{H_{total}}$ relationship depends on the gas to dust ratio, a 
quantity which is not only poorly constrained but can also vary by orders of 
magnitude for different astrophysical environments. So whilst the results of 
Figure \ref{figure:models} may only be valid beyond a certain extinction for 
our given A$_{v}$--N$_{H_{total}}$ relation, the results become valid at lower 
A$_{v}$ as the dust to gas ratio falls (since the C{\sc i} column
density would increase for a given A$_v$). We also note that for model F2 the 
C$\leftrightarrow$CO transition is pushed much deeper to A$_{v}\sim8$ by the 
more intense radiation field. This probably invalidates our assumption of CO 
being optically thick and perfectly self--shielded by A$_{v}=2$. However, one
effect of this would be to increase the abundance of atomic carbon at 
intermediate depths and hence the optical depth of the C{\sc i} continuum 
shielding CO. 

Figure \ref{figure:shieldings} shows the range of enrichments achieved when 
using different levels of shielding in model F1. The final panel shows that as 
carbon becomes optically thick, the resulting enrichments are very similar 
with or without H$_{2}$ line shielding, even though CN is not fully shielded 
by the carbon ionization continuum.
More importantly, the first panel shows that in environments 
where carbon column densities are low and only H$_{2}$ line shielding occurs, 
significant enrichments in N$_{2}$, N$_{2}$H$^{+}$ and NH$_{3}$ are still 
possible. Qualitatively similar behaviour is also seen for models F2 and F3. 
The exception is CN, which in model F1 is seen to be coupled to the enrichment 
in atomic carbon through the formation channel:
\[ {\rm C + NO \rightarrow CN + O.}\] 
Since the H$_2$ photodissociation lines become optically thick at low 
extinctions (A$_{v}\sim 0.1$), the effects of H$_2$ mutual shielding will 
always be significant for A$_v>2$. 
We therefore expect the true chemical enrichment for A$_v>2$ to be some 
intermediate between these two plots; that is to say, the effects of mutual 
shielding by the H$_2$ photodissociation lines will always be present even if 
the carbon ionization continuum is optically thin.

\section{Conclusions and Discussion}
\label{sec:conclusions}

We have re-calculated the interstellar photodissociation and photoionization 
rates of a range of interstellar molecules, both for direct interstellar and 
for cosmic ray induced radiation fields. 
We have also calculated direct photorates which include the effect of partial 
blocking of the incident radiation field by the carbon ionization continuum.

The specific quantitative results that can be drawn from our model are 
clearly limited by its simplicity. However, using approximations for shielding 
of the incident radiation field deep in PDRs by molecular hydrogen and atomic 
carbon, we present the result of significant (more than an order of 
magnitude) enrichments in the nitrogen-bearing species N$_{2}$, 
N$_{2}$H$^{+}$, NH$_{3}$ and CN. These results are seen for a range of 
physical parameters, including density, radiation field intensity and carbon 
column density, but should be most prominent in dust poor environments. These 
species are all important observationally in inferring various characteristics 
of the PDRs in which they are detected. For example, \citet{Bea:2009} have 
used PDR models to show that CN traces the density and relative nitrogen 
abundance of PDR dominated galaxies. First results from the Herschel Space 
Observatory towards W31C have shown nitrogen hydride (NH$_{3}$, NH$_{2}$ and 
NH) abundances that cannot be explained using simple gas--grain or PDR chemical 
modelling \citep{Pea:2010}. Other nitrogen-bearing species 
such as HCN and HNC, also believed to trace dense molecular gas, show smaller 
($<0.5$ magnitudes), correlated enrichments. They are often used in both 
galactic (e.g. \citealt{Robertsea:2011}) and extragalactic (e.g. 
\citealt{Kea:2011}) environments to classify the gas chemistry as photon 
dominated or not. It is clear from our work that the shielding by H$_{2}$ and 
C have a large impact on the abundances of such nitrogen-bearing species and 
both mechanisms need to be included in future PDR models.

Current ongoing studies suggest these mechanisms have only a limited effect on 
column densities in high extinction clouds and cannot explain, for example, 
the high column densities of nitrogen hydrides seen in the interstellar 
medium towards W31C \citep{Pea:2010}. In future work we will apply the 
shielding factors to more rigorous PDR models with self consistently calculated 
column densities to investigate the effects on the chemistry in regions of 
lower visual extinction (A$_v<2$). This is particularly important given that 
our observed enrichments appear more pronounced at lower extinctions where 
molecular shielding by H$_2$ becomes more significant relative to dust 
extinction. It would also be useful to do a thorough re-analysis of the cosmic 
ray induced photodissociation rate of CO, which is the dominant source of 
atomic carbon in regions of high extinction (A$_v>4-5$).

\section*{Acknowledgements}

We thank the anonymous referee for their commentary on the paper that led to 
significant improvements in its clarity.
RR acknowledges the financial support of the Science and Technology Facilities 
Council via a postgraduate studentship.

\bibliographystyle{mn2e}
\bibliography{ref}

\end{document}